\documentclass[letterpaper]{article}
\usepackage{isea}
\usepackage[pdftex]{graphicx}
\usepackage{times}
\usepackage{helvet}
\usepackage{courier}
\usepackage[numbers]{natbib}
 
\PassOptionsToPackage{hyphens}{url}\usepackage[colorlinks=false]{hyperref} 

\usepackage{soul} 

\usepackage{xcolor}

\usepackage{float}

\title{Rediscovering Korea's Ancient Skies: An Immersive, Interactive 3D Map of Traditional Korean Constellations in the Milky Way}
\author{Sung-A Jang$^1$, Benjamin L'Huillier$^2$ \\
$^1$Human Computer Interaction Institute, Carnegie Mellon University,
5000 Forbes Ave, Pittsburgh, PA 15213, USA,
sunga@cmu.edu\\
$^2$Korea Astronomy \& Space Science Institute,
776 Daedeok daero, Yuseong-gu, Daejeon 34055, Korea, benjamin@kasi.re.kr
\newline
\newline
}

\date{January 2019}

\begin{document}
\maketitle

\begin{abstract}
In this work, we visualized Korea's traditional constellations within an interactive 3D star map we created of the Milky Way. Unlike virtual planetariums based on celestial star coordinates from Earth's viewpoint, our visualization enables people to experience and interact with Korean constellation forms and its constituent stars in 3D space, and appreciate their historical, cultural significance from a contemporary perspective. 
Our interactive constellation map is based on the most detailed and accurate information on the stars in our Galaxy to date, and combines our expanding scientific understanding of the stars with contextual information reflecting Korea's unique astronomical culture and heritage.

\end{abstract}

\keywords{Keywords}
{
\footnotesize
Immersive Data Visualization, Virtual Reality, Korean Constellations, Star Maps, Astronomy
}

\section{Introduction}

Korea has a rich astronomical culture and history, based on a deeply rooted belief that celestial and terrestrial events are inextricably linked, and that the skies essentially hold the secrets to the Universe \cite{cha}. The considerable attention ancient Koreans paid to observing the skies is apparent in the detailed historical records of astronomical observations and interpretations, which reveal how integral and critical the skies were to their worldview \cite{park08}. 
Much of this is lost, however, in how 
Koreans see the stars today. The constellation shapes and stories Koreans have grown most familiar with are predominantly from Western cultures. The rich tapestry of Western celestial myths and starlore lives on to this day, in how we casually identify the stars and through popular superstitions. We aim to challenge this status quo by introducing different cultural perspectives to our modern star charts. Our goal is to encourage people to explore and enjoy the cutting-edge astronomical advances that enable new, exciting discoveries about the stars that populate our Galaxy, in tandem with the rich contextual information of the diverse ways in which we humans historically perceived and made sense of celestial movements and events, and the intricate systems we developed for interpreting them across different cultures and civilizations over centuries. 

Our specific focus was on visualizing Korea's ancient constellations and embedding its information within the interactive 3D star map we had created of our Galaxy. 
The purpose was twofold: (1) to encourage a rediscovery of Korea's own historical view of the skies, which has been woefully neglected and buried under prevailing Western conventions, and symbolically connect Korea's astronomical traditions and culture with current advances in modern astronomy; and (2) to show how much our connection with and knowledge of the skies have evolved since those ancient times. Interactive 3D visualization in virtual reality (VR) opens doors to new creative possibilities. By encouraging users to fly to anywhere in the Galaxy space inside the visualization, we want people physically experience how the constellation shapes become almost instantly unrecognizable when one breaks away from fixed viewpoint from Earth, and how two connected stars that appear close to each other may actually be separated by a vast distance in reality. These scientific truths are what ground our perception, and help us appreciate our rich cultural heritage in astronomy from a critical perspective. 

The project's overarching aim is to inspire people to rediscover the depth of history and culture hidden in the stars, and to explore the layers of meaning associated with the patterns and movement of celestial bodies exuding light, which guided our ancestors through the long dark nights.  

\section{Related Work}

The open software Stellarium\footnote{\url{https://stellarium.org/}} is a virtual planetarium for desktops that enables amateur astronomers to simulate the sky at a desired time and position. It has 600,000 stars, and allows users to turn on constellations from different cultures, including Korean. Their planetarium visualization of the sky, however, is based on 2D coordinates of the stars and is only capable of showing how the stars and constellations appear from locations on Earth. There are constellation labels in English available for viewing, but users cannot interact with the constellations to learn more about their underlying stories and cultural significance. The data for the Korean constellations is also solely based on the open contributions of a Korean Stellarium user made over a decade ago, whose sources are not identified and unverifiable. 

Figures in the Sky\footnote{\url{http://www.datasketch.es/may/code/nadieh/}} is a data visualization on the web by designer and astronomer Nadieh Bremer which compares the constellation shapes of different sky cultures. It features a selection of famous stars and visualizes the different constellation shapes they form depending on the culture. Her work also includes a full sky view of Korean constellation shapes, but it only appears as a static 360-degree panoramic view and we cannot interact with the shapes for additional information. The data used for the Korean constellations is also from Stellarium and hence unreliable. 

Star Chart\footnote{\url{https://www.oculus.com/experiences/rift/877457905696954}} is a VR planetarium app that provides an immersive real-time simulation of the night sky. The app allows users to view the entire sky and access information on constellations and individual stars. Despite the immersive view, however, the stars visualized only convey 2D information from the Earth's point of view, and only show the standard Western constellations. StarTracker VR\footnote{\url{https://www.oculus.com/experiences/gear-vr/1438854922813902}} also maps the star field onto a sphere surface and allows users to view the stars from outside the sphere. This, however, is also based on 2D data and falsely conveys the notion that the stars are all an equidistance from Earth. 

The European Space Agency (ESA) has released a full-sky interactive visualization of the stars\footnote{\url{http://sci.esa.int/gaia/60224-parallax-and-proper-motion-on-the-sky/}} based on their Gaia mission data, which shows an accelerated view of how stars move with time, from Earth's point of view. Western constellation lines are visualized, for the primary purpose of effectively showing the movement of the stars. Gaia VR\footnote{\url{http://sci.esa.int/gaia/60036-gaia-data-release-2-virtual-reality-resources/}} and Gaia Sky\footnote{\url{https://zah.uni-heidelberg.de/institutes/ari/gaia/outreach/gaiasky/}} are both 3D visualizations of the stars based on Gaia's data releases, and allow users to view Western constellation lines in 3D space. Neither visualization, however, allows users to explore additional information on the constellations nor includes constellations from other cultures.

\section{Data Collection \& Curation}

\subsection{\textit{Cheonsang-yeolcha-bunya-jido}} 
Made in the early \textit{Joseon} dynasty (1395) based on an ink rubbing of an ancient stele originating from the \textit{Goguryeo} kingdomm(37 BC--668 AD) according to its epitaph, the \textit{Cheonsang-yeolcha-bunya-jido} (“a planisphere chart of different celestial zones \cite{cha},” hereafter CS-map) is a monumental historic artifact representative of the East Asia's early advances in astronomical knowledge and tradition \cite{park08, yang12, ahn15}. A celebrated national treasure, the CS-map is a full-scale star chart that shows 1,467 stars forming 283 constellations \cite{cha}, a considerable proportion of which substantially differ from those in Chinese maps of a similar era \cite{yongbok, yang12}. The constellations are arranged according to the traditional East Asian constellation system (3 \textit{won} 28 \textit{su}), which divides the observable sky into 3 circles and 28 star groups. The 28 \textit{su} (lunar lodges) are constellations that appear in the celestial equator and mark the passage of the Moon in a sidereal month \cite{Stephenson94}, and they determine the boundaries which divide the 28 star groups \cite{ilgwon08}. 

\subsection{Building the Korean Star \& Constellation Database} 


There is unfortunately no digital database of the CS-map that we could reliably use to cross-match and identify modern catalog stars that form the constellations. The only pertinent data openly available online is the data used in Stellarium, which is allegedly based on the CS-map but unverifiable. This dearth of usable data prompted us to build our own digital database of Korean traditional constellations and stars, which consolidated information from reputable, validated academic sources detailed below to identify core constellation stars and extract their 3D position and magnitude\footnote{An astronomical unit proportional to the logarithm of the luminosity.} from modern catalogs via cross-analysis and verification.


\begin{figure}[h]
\includegraphics[width=3.31in]{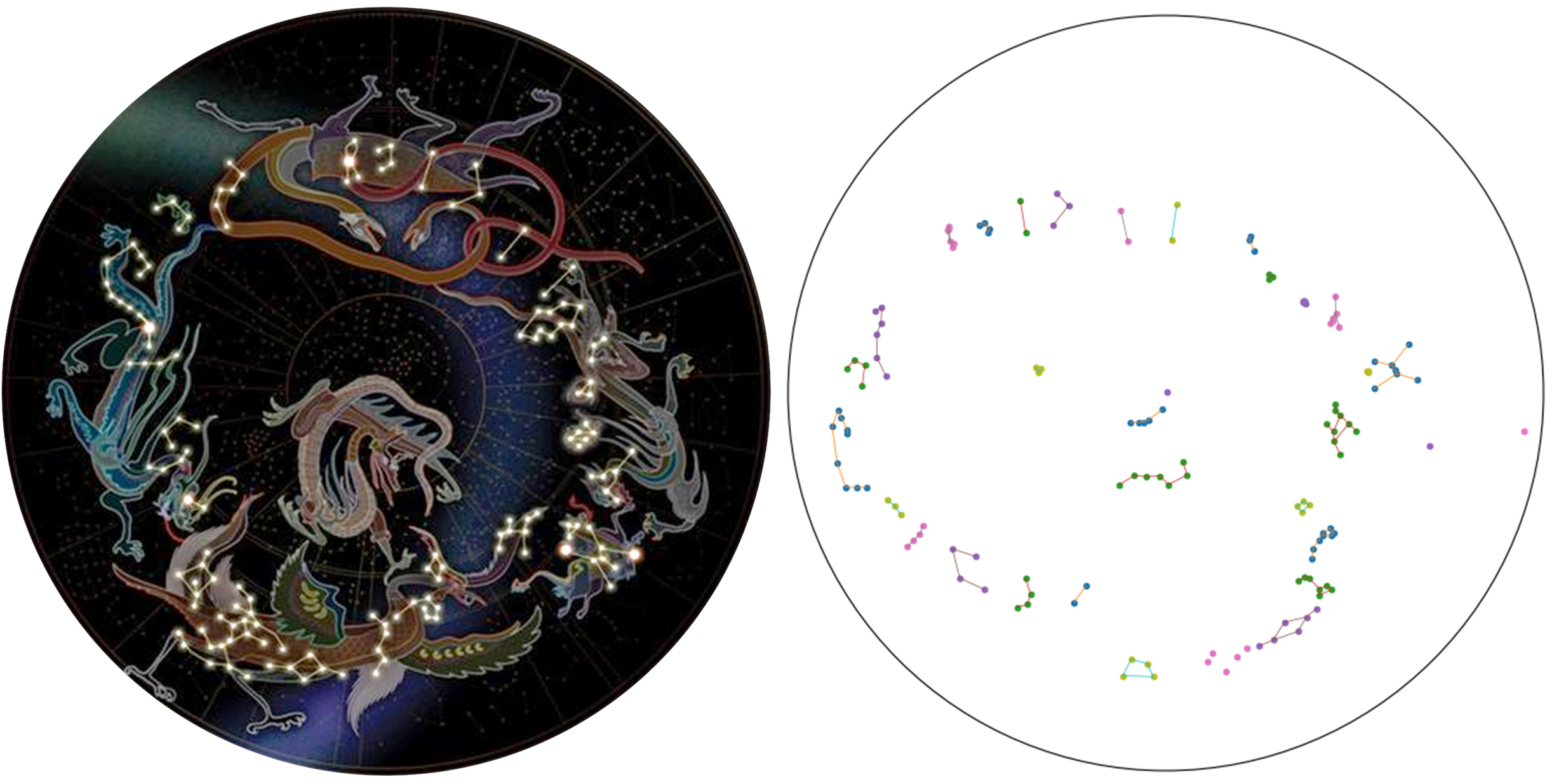}
\caption{\label{fig:2dverif}Visual verification via 2D and 3D mapping of compiled data. (Left image: \copyright Hong-Jin Yang)}
\end{figure}

Our starting point was Park's 1998 research paper on the CS-map \cite{park98}, which through rigorous computational analysis succeeded in identifying 375 of the 1467 stars that constitute the CS-map. Since approximately 3 quarters of the stars in the map were missing, we decided to constrain our initial visualization to the 28 main constellations that represented each \textit{su}, in addition to a selective number of culturally significant constellations that lay beyond this scope such as the \textit{Bukdu-Chilseong} (Seven Stars of the Northern Dipper), which corresponds to the Big Dipper asterism in Western astronomy and plays a pivotal part in East Asian mythology and starlore \cite{ilgwon08}.

Various supplementary historical astronomy sources \cite{AEEA, ahn15, yangdigi} were used as reference to complement Park's findings and fill in critical gaps in data as much as possible. Edge tables for each constellation were manually created in order to visualize the constellation forms. The identified constellations and stars were then visualized in 2D using modern catalog coordinates and compared with the CS-map for visual verification (Figure~\ref{fig:2dverif}).

Our ancient Korean constellations and stars database includes the traditional Korean names and meaning of each constellation and star, based on Ahn's book on unique Korean constellations \cite{ahnbook} and King Sejong's court astronomer Lee Sunji's classic astronomical text \textit{Cheonmun-Ryucho} from the 15th century \cite{sunji}. The database also details the stars' Western names, the Western constellations they respectively belonged to, and scientific information such as their exact magnitudes and Hipparcos IDs.

We used ESA Gaia mission's first data release (Gaia DR1\footnote{\url{https://gea.esac.esa.int/archive/}}, released in September 2016) as our primary star data source, since it was the most up-to-date and comprehensive star map available at the time and groundbreaking in its unprecedented coverage and accuracy. 
The positional information of the stars in the Gaia catalog is three-dimensional, in a spherical coordinate system: two angles (right ascension and declination) describing their position in the sky, and parallax. Parallax refers to the apparent motion in the sky of an object as seen from two opposite points of the Earth revolution around the Sun, and is used to infer the distance to ``close'' objects, i.e., within the Milky Way. 
Our data fused the portion of Gaia DR1 which provided the 3D position and brightness of 2 million stars, with the data of 24,320 bright stars from the Hipparcos catalog\footnote{\url{https://www.cosmos.esa.int/web/hipparcos/catalogues}} that were not included in Gaia DR1. 
From Ref.~\cite{park98}, we obtained the Yale Bright Star Catalog IDs of the identified stars on the CS-map, and used them to retrieve their respective IDs in the Hipparcos catalogue via the Hipparcos-Yale-Gliese (HYG) compilation.  
We then obtained their astrophysical information (2D/3D positions, magnitude) from the Hipparcos and Gaia catalogues. 
We derived the Cartesian coordinates of the stars from the angular coordinates and parallax data, and converted their apparent magnitude (as seen from Earth) to absolute (as seen from a distance of 10 parsecs, or 30 light-years) for visualization in Unity3D. 
We removed all stars with non-positive parallaxes.

\section{System Design}

\begin{figure}[h]
\includegraphics[width=3.31in]{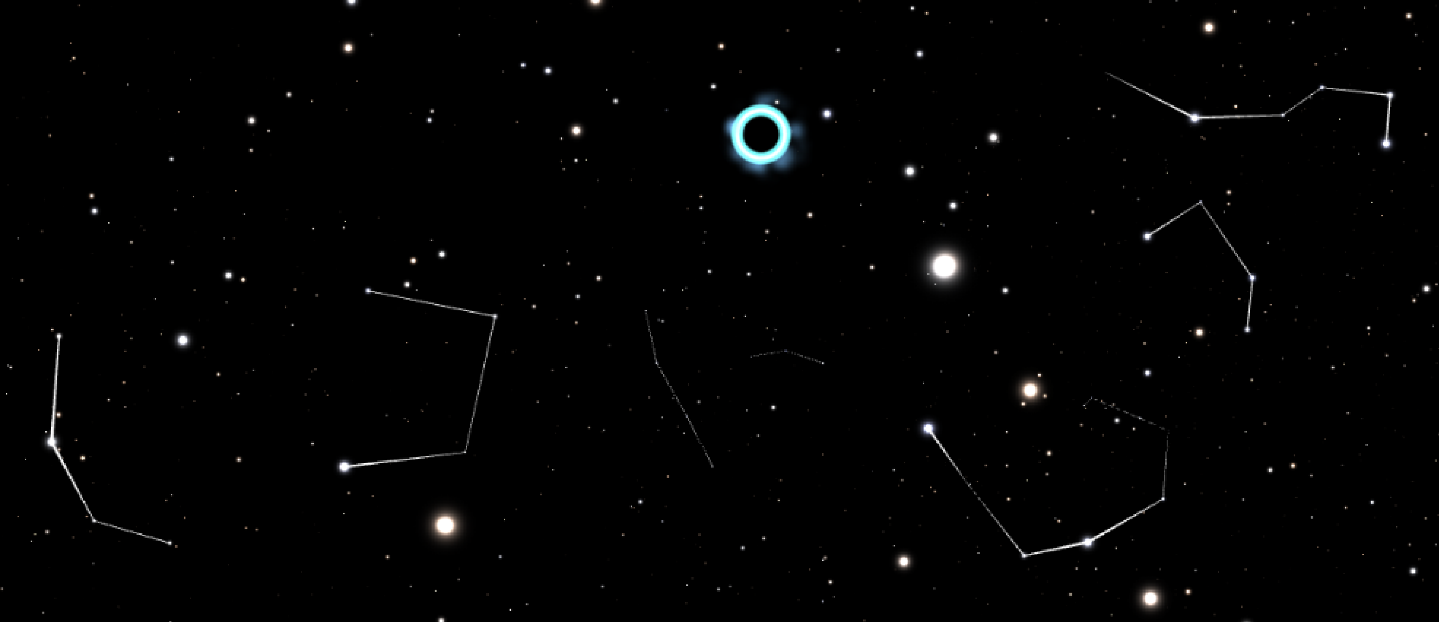}
\caption{\label{fig:kconstview} A view of the lunar lodges from the perspective of Earth.}
\end{figure}

\begin{figure}[h]
\includegraphics[width=3.31in]{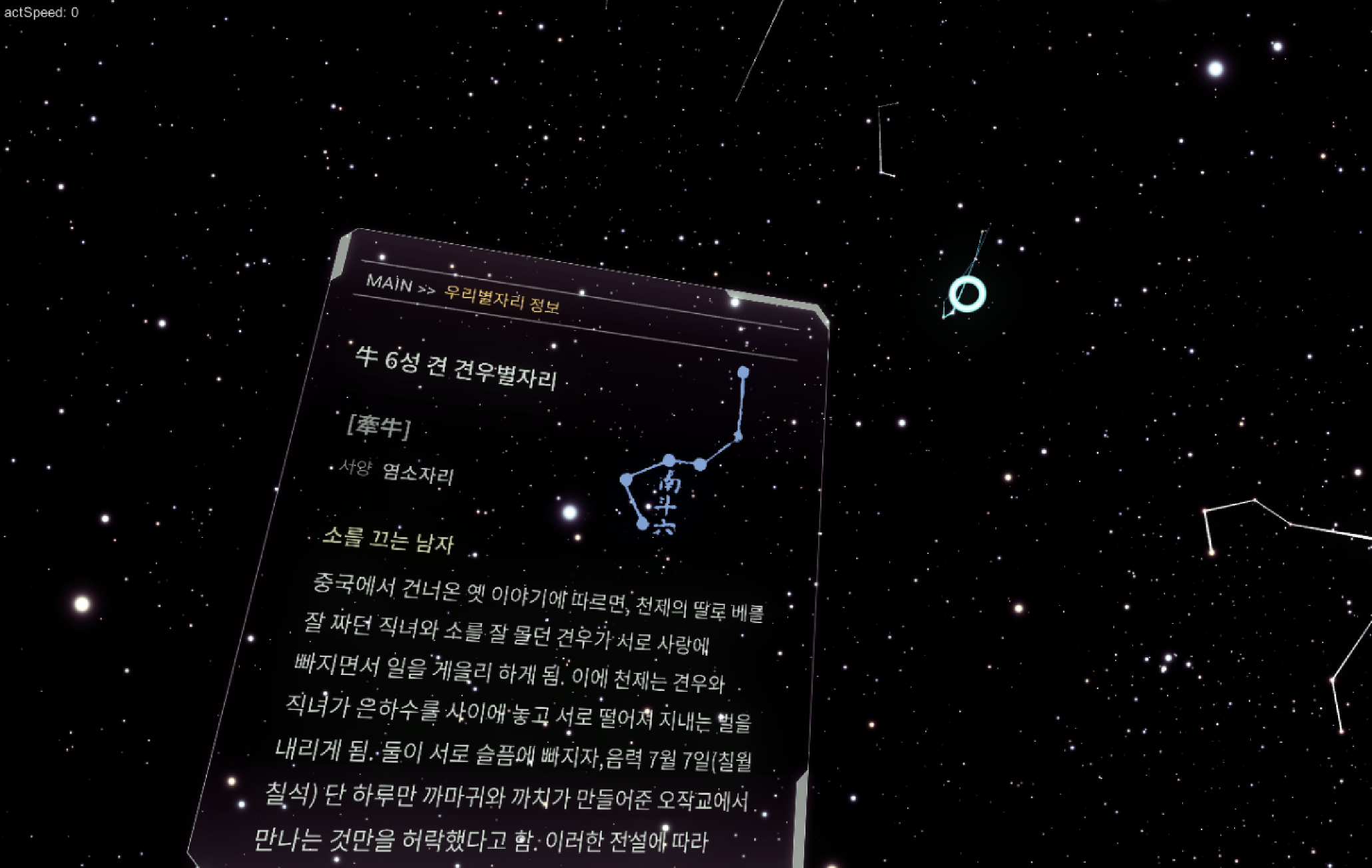}
\caption{\label{fig:kyeonwoo} The selected constellation's information panel appears in world space, facing the user from a comfortable viewing distance.}
\end{figure}

The system presents an immersive data-driven experience of Korean constellations and stars in our Galaxy. 25 of the 28 \textit{su} are visualized, along with a select number of important constellations inside the circle of perpetual visibility (Figure~\ref{fig:kconstview}), which encircles an area of the sky that could be observed in all seasons. Users are free to roam anywhere in the Galaxy and view the constellations from constantly changing perspectives. Specific information on every constellation and its stars are accessible by pointing the reticle in its direction and clicking the trigger button with the touch controller (Figure~\ref{fig:kyeonwoo}). 

\begin{figure}[h]
\includegraphics[width=3.31in]{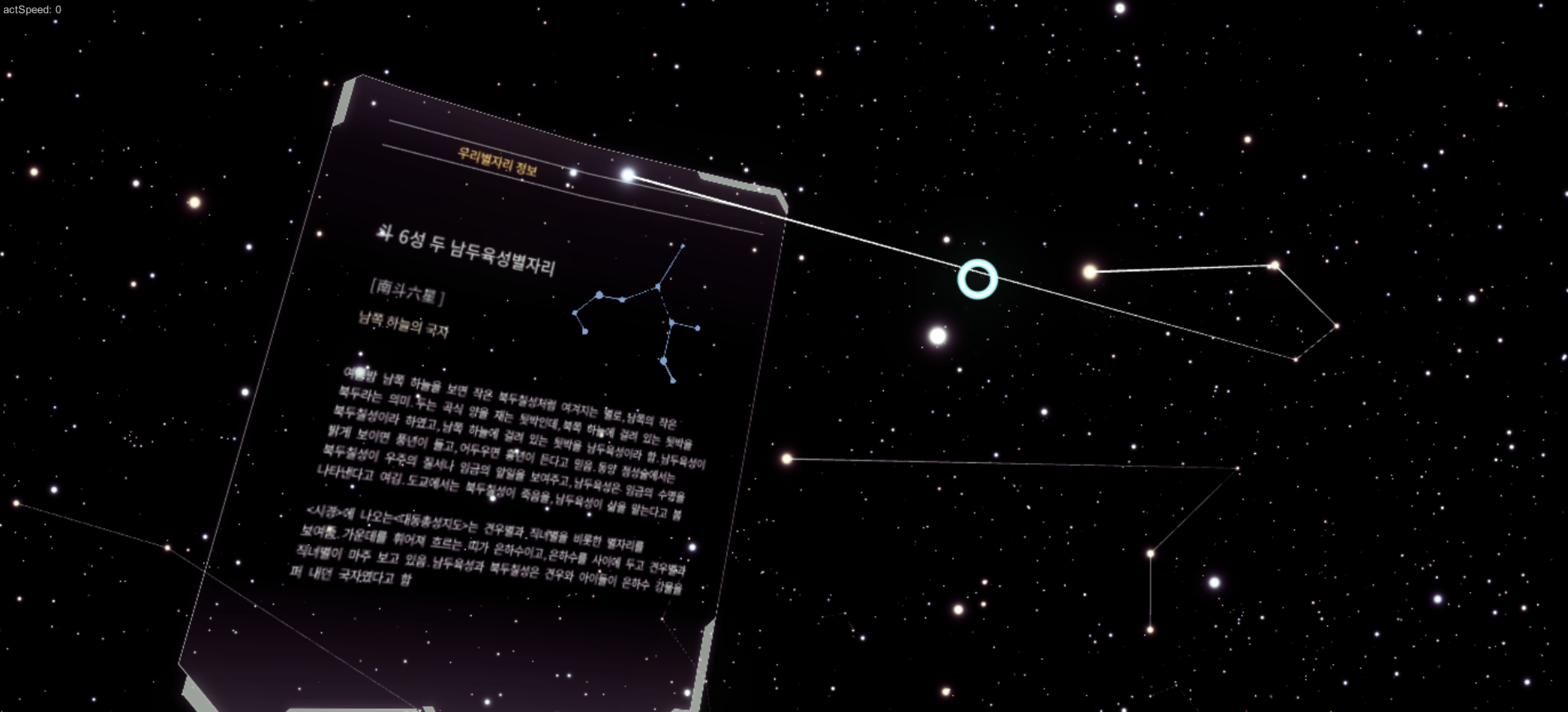}
\caption{\label{fig:namduview} The Southern Dipper constellation shown is drastically different from its iconic shape from Earth displayed in the info panel.}
\end{figure}

\begin{figure}[h]
\includegraphics[width=3.31in]{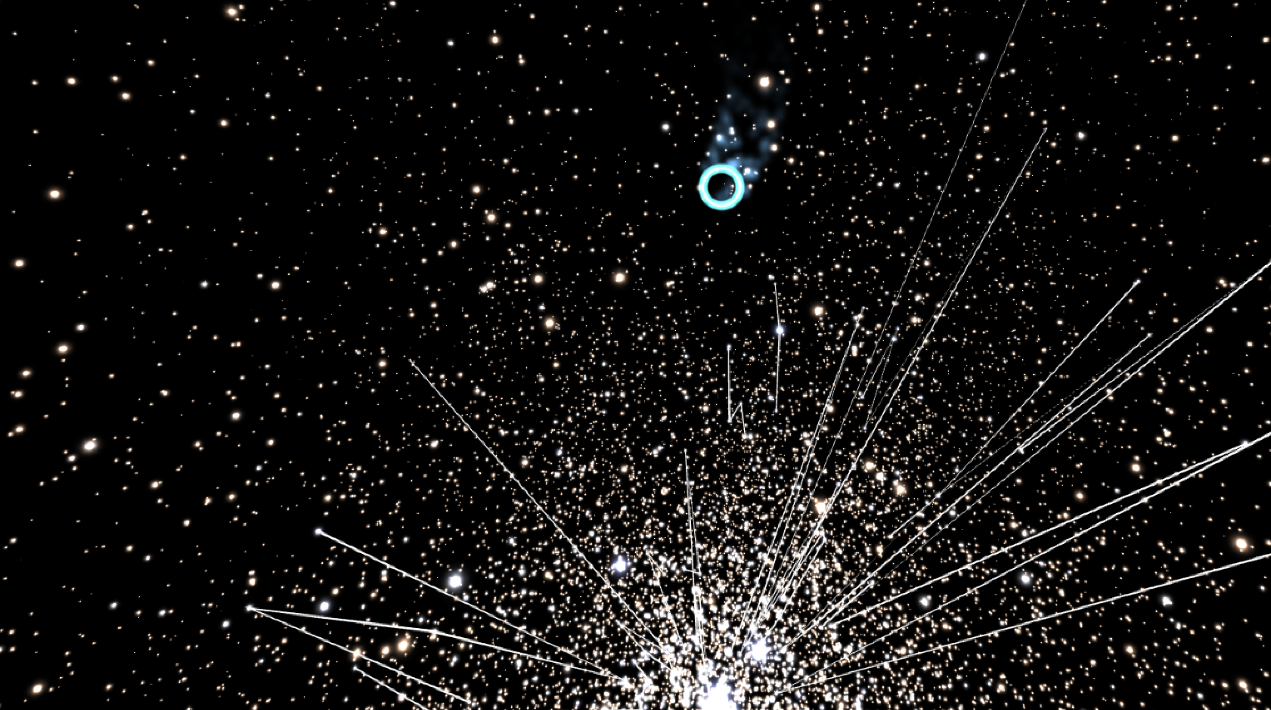}
\caption{\label{fig:distantview} The view of constellations and constituent stars from a point of view greatly removed from the Solar System.}
\end{figure}


The system was implemented in Unity3D and designed to work with the Oculus Consumer Version 1 VR headset and Oculus Touch controllers. The experience starts from the position of Earth and our Solar System. The left controller thumbstick controls the 3D navigation in space, and the navigation speed changes according to the degree to which the stick is pushed. Users can choose to move forward in the direction of their gaze, or set to absolute directions. The right controller's thumbstick controls the rotation of their view, and enables them to zoom in and out of our Galaxy (Figure~\ref{fig:distantview}). 

We intentionally designed a world-space interface instead of a classic heads-up display (HUD) view for toggling view options and displaying the information of selected constellations. As opposed to information being presented on a helmet-like display that moves with the users, the display panels dynamically appear in their world space out of thin air, facing the viewer from a comfortable distance, maximizing the users' sense of space. The displayed content is swiftly replaced when users remain in place and change their selections. Even when users are on the move, open panels stay fixed in world space where they first emerged until users close them or make a selection from a different point in space, at which point any previously open panels are instantly cleared. 

\begin{figure}[h]
\includegraphics[width=3.31in]{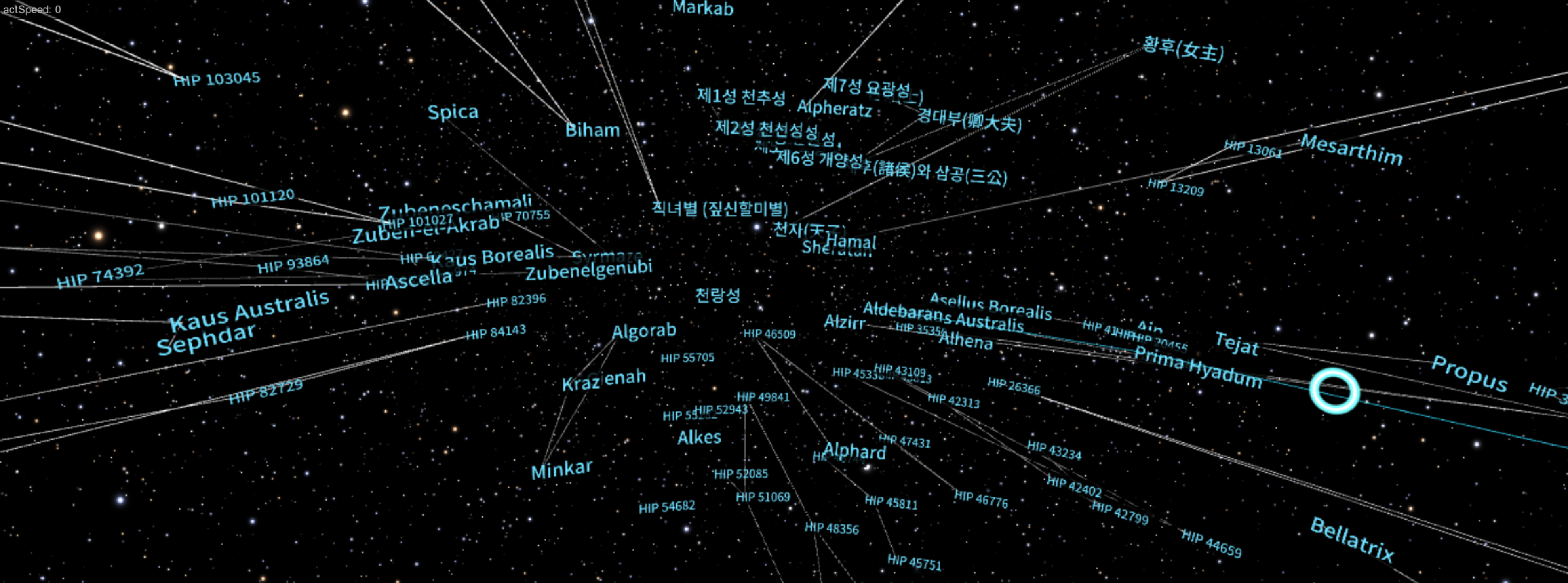}
\caption{\label{fig:textview} When the star labels are turned on, it prioritizes unique Korean names over Western ones when available. Hipparcos IDs are displayed for stars without distinct names.}
\end{figure}

Some constellations such as the \textit{Buk-du} (Northern Dipper) have important constituent stars that can be individually selected for detailed information. 
To access this information, users must hold the trigger button and release it when the reticle is hovering over the star. Users can also fly away from our Galaxy to gain a distant view, promptly return to their starting point (Earth), and toggle star and constellation names on and off in the view settings (Figure~\ref{fig:textview}). 

\section{Conclusion}
The system enables users to move freely in any direction they want in variable speeds faster than the speed of light, and explore constellations and starscape in motion. Freed from physical constraints that limit their views to a fixed position on Earth, users can experience how dramatically the constellation shapes and star configurations change depending on their viewpoint in space (Figure~\ref{fig:namduview}).
Imbuing the stars in the Galaxy with cultural history and context makes the Galactic data itself more engaging and approachable to the public. This project is part of a broader effort to connect recent astronomical advances with Korea's historical astronomy, and enliven interest in the new discoveries made as well as Korea's rich astronomical tradition and cultural heritage. We also strive through these endeavors to broaden access to rich sources of open data and transform it into public knowledge and appreciation. 

\section{Acknowledgements}
We thank Goeun Choi, Changbom Park, Cris Sabiu, and Hong-Jin Yang for their help in gaining access to CS-map data and much constructive feedback. This project was funded by the Korea Culture Technology Institute in Gwangju, Korea. 

\bibliographystyle{isea}
\bibliography{kconst}

\section{Authors Biographies}
Sung-A Jang is an artist and HCI researcher working in the intersection of art, science and technology. Benjamin L'Huillier is a cosmologist and astrophysicist who seeks to understand the content of the universe and the laws that govern its evolution.
\end{document}